\documentstyle [12pt,epsf]{article}
\newcommand{\beq}{\begin{equation}}
\newcommand{\eeq}{\end{equation}}
\newcommand{\bea}{\begin{eqnarray}}
\newcommand{\eea}{\end{eqnarray}}
\newcommand{\beqn}{\begin{eqnarray}}
\newcommand{\eeqn}{\end{eqnarray}}
\newcommand{\dd}{\partial}

\newcommand{\phib}{\bar{\phi}}

\newcommand{\etab}{\bar{\eta}}

\begin{document}
\baselineskip 7.5 mm

\def\thefootnote{\fnsymbol{footnote}}

\begin{flushright}
\begin{tabular}{l}
UPR-679-T \\
NSF-ITP-95-109 \\
hep-ph/9509275 \\
September, 1995
\end{tabular}
\end{flushright}

\vspace{20mm}

\begin{center}

{\Large \bf Supersymmetry in the false vacuum.
}
\\

\vspace{20mm}

\setcounter{footnote}{0}

Alexander Kusenko\footnote{ email address: sasha@langacker.hep.upenn.edu}
\vspace{4mm}\\
Department of Physics and Astronomy \\
University of Pennsylvania, Philadelphia, PA 19104-6396 \\
and \\
Institute for Theoretical Physics \\
University of California, Santa Barbara, CA 93106-4030 \\

\vspace{20mm}

{\bf Abstract}
\end{center}

A metastable state, or a false vacuum, is not an eigenstate of the
Hamiltonian in quantum field theory.  Its energy density has a non-zero
imaginary part equal to its decay width.  Therefore, supersymmetry cannot
be exact in the false vacuum.  We calculate the size of this effect using
the path integral approach.

\vfill

\pagestyle{empty}

\pagebreak

\pagestyle{plain}
\pagenumbering{arabic}
\renewcommand{\thefootnote}{\arabic{footnote}}
\setcounter{footnote}{0}

\pagestyle{plain}

The Universe is believed to have arrived at its present state via a series
of phase transitions, each of which involved a decay of some metastable
quantum state.  A good understanding of these transitions is essential for
explaining the physical world at present.  In particular, the baryon
asymmetry of the Universe, dark matter, and many other cosmological and
experimental data may owe their existence to the physics of the early
Universe.  Although the experimental evidence for supersymmetry has yet to
be discovered, theories with low-energy supersymmetry provide an attractive
explanation for the hierarchy of scales and are widely considered viable.
It is of interest, therefore, to understand the effects of
supersymmetry on the decay of a metastable state.

It is well-known that supersymmetry is broken at non-zero temperature
\cite{lr}.  There is, however, an additional effect, a generic feature of
the false vacuum which is due to its instability {\it per se}, which can
contribute to SUSY breaking\footnote{In application to the false vacuum,
the term SUSY {\it breaking} is somewhat a misnomer because supersymmetry
is not really broken.  The unstable state that is considered to be a
``vacuum''  is simply not a part of the spectrum and the corresponding Fock
space is ill-defined.  Nevertheless, for all intents and purposes SUSY will
appear broken to an observer in the false vacuum, which justifies the use
of the term.} and be equally or more important than the
finite-temperature effects.

Unlike the true vacuum, $|0 \rangle$, a false vacuum $|0' \rangle$
is not an eigenstate of the Hamiltonian $H$ \cite{c}.  It's
decay width per unit volume, $(\Gamma/{\sf V})$, is equal to the imaginary
part of the energy density,  $Im \langle 0' |H| 0'\rangle$, which
must be non-zero if the false vacuum is to decay.  On the other hand, by
the usual argument, exact supersymmetry
cannot be realized unless the vacuum energy is zero.  The latter
requirement implies that both the real and imaginary parts of
the energy density must vanish.  Indeed, the
anticommutation relation between the supersymmetry generators $Q_\alpha $

\beq
\{ Q_\alpha, Q_\beta^{\dag} \} = 2 \sigma^{\mu}_{\alpha \beta} P_{\mu},
\eeq
together with the statement of unbroken supersymmetry,

\beq
Q_\alpha |0' \rangle=Q_\beta^{\dag} |0' \rangle =0,
\eeq
yield immediately that

\beq
\langle 0' | P_{\mu} | 0' \rangle = 0 \ \Longrightarrow \
Im \langle 0'| P_{\mu} |0' \rangle =0 \Longrightarrow \
\Gamma/{\sf V}=0,
\eeq
in contradiction with the assumption that $|0' \rangle $ is a metastable
state and thus has a non-zero decay width.

This simple line of reasoning shows that supersymmetry cannot be exact in
any metastable state and that the vacuum in which SUSY is unbroken must be
stable.  The latter can serve as an independent argument for the stability of
the supergravity vacuum with vanishing cosmological constant even in the
presence of lower lying minima in the potential, a fact established in
Ref. \cite{w} by other means.

\begin{figure}
\setlength{\epsfxsize}{3.5in}
\setlength{\epsfysize}{2.5in}
  \centerline{\epsfbox{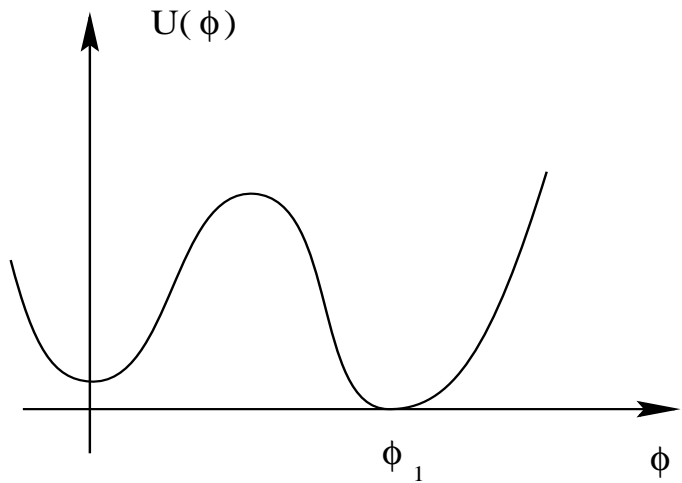}}
\caption{The false and the true vacua.
}
\label{fig1}
\end{figure}

It is clear that the size of the SUSY breaking effects in the false vacuum
is of order $\Gamma/{\sf V}$.  Below we will show that this is, in fact, the
case and will compute the mass splitting $\Delta m^2= m_{_B}^2-m_{_F}^2 \sim
\langle F \rangle$ for the bosonic and fermionic components of a chiral
superfield which arises due to non-perturbative effects related to false
vacuum instability.  In general, there are other contributions to
$\Delta m^2$ which can be computed in perturbation theory.  For example, if
the scalar potential is of the form Fig. \ref{fig1}, and if supersymmetry
is unbroken in the true minimum, then there will be a contribution
$(\Delta m^2)_{_R}$ to $\Delta m^2$ from the real part of the energy density
due to spontaneous or explicit (soft) SUSY breaking at $\phi=0$.
In realistic models one can assume $(\Delta m^2)_{_R}\sim M^2_{_{SUSY}}$.
At finite temperature supersymmetry is always broken \cite{lr} which gives
rise to a temperature-dependent contribution $(\Delta m^2)_{_T}$.
So, in general,

\beq
\langle F \rangle \sim \Delta m^2=(\Delta m^2)_{_R}+(\Delta
m^2)_{_T}+(\Delta m^2)_{_I}
\label{deltam2}
\eeq
where $(\Delta m^2)_{_I}$ is the non-perturbative piece induced by
the false vacuum instability.  The relative magnitudes of the terms in
(\ref{deltam2}) are model-dependent.  However, one can always construct a
model in which the barrier separating the false vacuum from the true is
sufficiently small, so that the last term in equation (\ref{deltam2})
is arbitrarily large, while the first two terms are much smaller.

We see that the supersymmetry breaking due to the instability of the false
vacuum will dominate over that due to finite-temperature and other
effects for a metastable state that is sufficiently short-lived,
or, in other words, if $ \Gamma/{\sf V} \gg M_{_{SUSY}}^4$ and
$ \Gamma/{\sf V} \gg T^4 $.
In what follows we will neglect the $(\Delta m^2)_{_R}$ and $(\Delta
m^2)_{_T}$  which can be calculated perturbatively in each given model and
will concentrate on $(\Delta m^2)_{_I}$.

We now evaluate the mass splitting of the fermion $\psi$ and
boson $\phi$ components in a model involving a single chiral superfield
$\Phi=\{\phi, \psi, F \}$ with the scalar potential that has two
non-degenerate local minima (Fig. \ref{fig1}).

In the path integral formulation a vacuum expectation value
of some function $f(\phi)$ of the dynamical field can be written as

\beq
\langle f(\phi) \rangle = \lim_{J \rightarrow 0}
\frac{\delta}{\delta J(x_0)} \ \ln \ \int [d \phi]  \
\exp(-\frac{1}{\hbar} S[\vec{\phi}]+\int J f(\phi)),
\label{int}
\eeq
where $S[\vec{\phi}]$ is the Euclidean action and $J(x)$ is the source.

The non-perturbative contribution to $f(\phi)$ can be evaluated in the
semiclassical limit $\hbar \rightarrow 0$ using the saddle point
approximation (see, {\it e.\,g.}, \cite{c,cc}).  The idea of the method is
that one can expand the path integral around the finite-action solutions
which are saddle points of the Euclidean action and are, therefore,
determined by the variational equation

\beq
\delta S = 0.
\label{dS}
\eeq

For a fermion $\psi(x)$, the Euler-Lagrange equation (Euclidean equation
of motion) corresponding to equation (\ref{dS}) is a first-order
differential equation.  The boundary conditions
$\psi(x\rightarrow \infty)=0$ are necessary to ensure the finiteness of
the action.  The only solution $\psi(x)$ of equation (\ref{dS}) satisfying
these boundary conditions is the trivial one $\psi(x) \equiv 0$.

In contrast to fermions, the bosonic equation of motion corresponding to
(\ref{dS}) is second-order and it allows for a non-trivial solution of
finite action, the so called ``bounce'' \cite{c}.  The bounce dominates the
path integral contribution to the imaginary part of the energy density,
which was calculated in Refs. \cite{c,cc,ak_g}.   More specifically, the
bounce for a scalar potential $U(\phi)$ (Fig. 1) is  a nontrivial
$O(4)$-symmetric \cite{c_comm} field configuration determined by the
equation\footnote{
If the potential depends on several scalar fields, then finding the bounce
numerically becomes a non-trivial task.  In this case, the method of
Ref. \cite{ak_i} is useful.
}

\begin{equation}
   \Delta \bar{\phi}(r)= \frac{\partial}{\partial \bar{\phi}}
U(\bar{\phi})
\label{bounce_eq}
\end{equation}
with the following boundary conditions:

\begin{equation}
\left \{ \begin{array}{l}
    (d/dr) \bar{\phi}(r)|_{r=0}=0 \\  \\
   \bar{\phi}(\infty)=0
        \end{array} \right.
\label{boundary}
\end{equation}
where $r=\sqrt{x^2}$.

The potential $U(\phi)$ can be written as

\beq
U(\phi)=\frac{m_0}{2} \phi^2 + \sum_{n>2} c_n \phi^n.
\label{V}
\eeq
and the boson mass, by definition, is

\bea
m_\phi^2 & = & \left \langle \left ( \frac{\dd^2 U(\phi)}{\dd \phi^2}
\right ) \right \rangle  \nonumber \\ & & \nonumber \\
  & = & \frac{\int \left ( \frac{\dd^2 U(\phi)}{\dd \phi^2}\right )
e^{-S[\phi]} [d\phi ]}{\int e^{-S[\phi]} [d\phi ]} \nonumber \\ & &
\nonumber \\
 & = & m_0^2 + \frac{\int f(\phi) e^{-S[\phi]} [d\phi ]}{
\int e^{-S[\phi]} [d\phi ]}
\label{dm}
\eea
where $f(\phi)=(\dd^2/\dd \phi^2) \sum_{n>2} c_n \phi^n $.  We note that
$f(0)=0$.

We now apply the saddle point method to calculate the quantity
$\Delta m^2=m_\phi^2 - m_0^2$.  First we will compute the normalization
factor

\beq
N=\int e^{-S[\phi]} [d\phi ]
\label{N}
\eeq
in the denominator of the second term in (\ref{dm}).
This part of the calculation will follow closely that of Refs. \cite{c,cc}.
For simplicity, we consider the case of one space-time dimension $t$;
generalization to $d=4$ is straightforward and will follow.

The bounce of equation (\ref{bounce_eq}) is centered at $t=0$ because
$\bar{\phi}'(0)=0$.  Let's consider a contribution from a
single bounce $\phi(t-t_j)$ centered at $t=t_j$, $-T/2<t_j<T/2$, where $T$
is some large quantity taken eventually to infinity.
A classical field configuration corresponding to $n$ widely separated
bounces will also be a saddle point of the integrand with the Euclidean
action $n S[\phib]$.  Therefore, we have to
sum the contributions of all such saddle points and integrate
over the positions of the bounces.

We are interested in the limit $T \rightarrow \infty$. The $n$ bounces
centered at $t_1, \ t_2, \  ... \ t_n$, where $-T/2 <t_n < ...<
t_2< t_1 < T/2$,  are separated by vast regions where the field is
nearly zero.  Therefore the determinant part of the path integral can be
written as a product

\begin{equation}
\sqrt{(\omega / \pi \hbar)} e^{-\omega T/2} K^n
\end{equation}
where $\omega=U''(0)$ and
$K$ is defined so as to yield the correct answer for $n=1$.

We must now integrate over the locations of the centers of the bounces.

\begin{equation}
\int_{-T/2}^{T/2} dt_1 \int_{-T/2}^{t_1} dt_2 \ ...\
\int_{-T/2}^{t_{n-1}}
dt_n = \frac{T^n}{n!}
\label{Tn}
\eeq

The summation of contributions with different $n$ gives:

\begin{eqnarray}
\sum_{n=1}^{\infty} (\frac{\omega}{\pi \hbar})^{1/2}
e^{-\omega T/2} K^n (e^{-S[\bar{\phi}]/\hbar})^n \frac{T^{n}}{n!} =
\nonumber \\ \nonumber \\
(\frac{\omega}{\pi \hbar})^{1/2} \ \exp(-\omega T/2+ K
e^{-S[\bar{\phi}]/\hbar} T)
\label{sum1}
\end{eqnarray}
This is the same as the corresponding result in \cite{cc}.  The quantity
$K$ is the correction to the energy which determines the transition
probability.  We will come back to it later.

Now we treat the numerator of the second term in (\ref{dm}).  This is
basically the same, except now the configuration which comprises $n$ widely
separated bounces, $ \phi=\sum_{j=1}^{n} \bar{\phi}(t_0-t_j)$,
will contribute

\begin{equation}
\sum_{j=1}^{n} f[\bar{\phi}(t_0-t_j)] \ e^{-n S[\bar{\phi}]/\hbar},
\end{equation}
where the property $f(0)=0$ is essential.

Like before, we have to integrate over the positions of the bounces:

\begin{equation}
\int_{-T/2}^{T/2} dt_1 \int_{-T/2}^{t_1} dt_2 \ ...\
\int_{-T/2}^{t_{n-1}}
dt_n  \sum_{j=1}^{n} f[\bar{\phi}(t_0-t_j)]
\end{equation}

To evaluate the integral, we exchange the order of
summation and integration.  Then we let each variable $t_i$ run over
the entire interval $[-T/2,T/2]$.  The latter change
would lead to a double counting of
the contributions differing by permutations of the identical bounces.  We
therefore divide by $(n!)$, the number of permutations of $n$ bounces:

\begin{eqnarray}
\sum_{j=1}^{n} \frac{1}{n!}
\int_{-T/2}^{T/2} dt_1 \int_{-T/2}^{T/2} dt_2 \ ...\ \int_{-T/2}^{T/2}
dt_n   \bar{\phi}(t_0-t_j) =  \nonumber \\ \nonumber \\
\frac{1}{n!} \sum_{j=1}^{n}
\int_{-T/2}^{T/2} dt_1 \ ... \ \{\int_{-T/2}^{T/2} dt_j
f[\bar{\phi}(t_0-t_j)] \} \ ... \ \int_{-T/2}^{T/2} dt_n
\end{eqnarray}
In the last expression, the integration over $t_j$ gives
(in the limit of large $T$) $\int_{-\infty}^{\infty} \bar{\phi}(\xi) d\xi$,
while each of the other integrals gives a factor of $T$.  The sum over $j$
has $n$ identical terms and is equal to $n$ times the value of each.
The factor $n$ then cancels with that of $n!$.  Altogether we get

\begin{equation}
\{ \int_{-\infty}^{\infty} f[\bar{\phi}(\xi)] d\xi \: \}  \
\frac{T^{n-1}}{(n-1)!}
\label{Tn-1}
\end{equation}
We notice in passing that the expression (\ref{Tn-1}) differs from
(\ref{Tn}) in two respects:  (1) there is a constant factor of
$\int f[\bar{\phi}(\xi)] d\xi$; (2) it has one less power of $T$.

And, finally, we sum over $n$:
The $n=0$ component will not
contribute; therefore, the sum starts from the $n=1$ term:

\begin{eqnarray}
\sum_{n=1}^{\infty} (\frac{\omega}{\pi \hbar})^{1/2}
\{\int_{-\infty}^{\infty} f[\bar{\phi}(\xi)] d\xi \}
e^{-\omega T/2} K^n (e^{-S[\bar{\phi}]/\hbar})^n \frac{T^{n-1}}{(n-1)!}=
\nonumber
\\ \nonumber
\\
\{ \int_{-\infty}^{\infty} f[\bar{\phi}(\xi)] d\xi \} \ \
K e^{-S[\bar{\phi}]/\hbar} \ (\frac{\omega}{\pi \hbar})^{1/2} \
\sum_{n=1}^{\infty}
e^{-\omega T/2} K^{n-1} (e^{-S[\bar{\phi}]/\hbar})^{n-1}
\frac{T^{n-1}}{(n-1)!} =
\nonumber \\ \nonumber \\
\{ \int_{-\infty}^{\infty} f[\bar{\phi}(\xi)] d\xi \} \ \
K e^{-S[\bar{\phi}]/\hbar} \
(\frac{\omega}{\pi \hbar})^{1/2} \ \exp(-\omega T/2+ K
e^{-S[\bar{\phi}]/\hbar} T)
\label{sum}
\end{eqnarray}

As we now divide expression (\ref{sum}) by (\ref{sum1}), we obtain a simple
result:

\beq
m_\phi^2=m_0^2+\Delta m^2= m_0^2+
\{ \int_{-\infty}^{\infty} f[\bar{\phi}(\xi)] d\xi \}
\ \  K e^{-S[\bar{\phi}]/\hbar}
\label{dmK}
\eeq
where the factor $K$ is that computed in \cite{cc}\footnote{
In applications, one is often interested in transitions between phases
with different symmetries.  In this case, some additional zero modes
appear in the determinant of equation (\ref{dmf}) due to the
Goldstone phenomenon.  The corresponding $N$ zero eigenvalues should then
be omitted from the  ``primed'' determinant in (\ref{dmf}) and an additional
factor  $\left [\int \etab^2(x) d^4x \right ]^{N/2}$ will appear in the
expression for $\Delta m^2$ \cite{ak_g}.
}:

\beq
|\Delta m^2| = \{ \int_{-\infty}^{\infty} f[\bar{\phi}(\xi)] d\xi \}
\left(\frac{S[\phib]}{2\pi \hbar} \right)^2
e^{-S[\phib]/\hbar}
\left | \frac{\det'[-\dd_\mu^2+U''(\phib)]}{\det[-\dd_\mu^2+U''(0)]}
\right |^{-1/2}
\times (1+O(\hbar))
\label{dmf}
\eeq
where $\det'$ stands for the determinant from which all zero eigenvalues
are omitted.

As was explained earlier, there is no analog of the bounce for fermions,
so that the correction (\ref{dmf}) applies to bosons only.

The above discussion dealt with the zero-temperature tunneling.  It is
straightforward to generalize the expression (\ref{dmf}) to the
finite-temperature case \cite{linde}.  The changes will amount to
replacing the Euclidean action of the bounce with the action of a
three-dimensional bounce $S_3$ divided by the temperature, and the
factor $(S/h)^2$ with $(S_3/h)^{3/2}/T^{1/2}$.

In summary, we have computed the mass splitting between the components of
a chiral superfield which is related to the instability of the false
vacuum.  If the metastable state decays sufficiently fast, this
effect may dominate over the SUSY breaking finite-temperature
effects, as well as the soft SUSY breaking terms of phenomenologically
acceptable size. We have also given an independent argument for the
supersymmetric vacuum stability in supergravity.

The author would like to thank P. Langacker for many stimulating
conversations and helpful comments.  We are also  grateful to M. Dine for
a useful discussion.  This work was supported by the U.~S.~Department
of Energy Contract No. DE-AC02-76-ERO-3071 as well as by the National
Science Foundation Grant No. PHY94-07194.

\end{document}